# PHOTON STATES FROM PROPAGATING COMPLEX ELECTROMAGNETIC FIELDS

D. Dragoman – Univ. Bucharest, Physics Dept., P.O. Box MG-11, 077125 Bucharest, Romania, e-mail: danieladragoman@yahoo.com

ABSTRACT

A wavefunction for single- and many-photon states is defined by associating photons with different momenta to different spectral and polarization components of the classical, generally complex, electromagnetic field that propagates in a definite direction. By scaling each spectral component of the classical field to the square root of the photon energy, the appropriately normalized photon wavefunction acquires the desired interpretation of probability density amplitude, in contradistinction to the Riemann-Silbertsein wavefunction that can be considered as the amplitude of the photon probability energy density.

1. INTRODUCTION

Despite claims that massless particles cannot be localized in space-time [1, 2], quantum optics has long struggled to define a quantum wavefunction for photons that would be in agreement with innumerable experiments revealing the possibility of photon localization. In particular, in any quantum electrodynamics textbook the quantum state of a many-photon system is easily obtained in terms of creation and annihilation operators using the mathematical similarity between the harmonic oscillator and the electromagnetic field. This theory explains many quantum optics experiments, although it is based on incongruities related to the equivalence of the photon, which is massless according to relativity theory, with a harmonic oscillator with unit mass, and to the introduction of the troublesome zero-point energy [3], which is supported by experimental results relating to the Casimir effect and Lamb shift, for example, but rejected by cosmological observations [4]. Even more controversies plague the attempts to introduce single- or few-photon quantum wavefunctions. All proposals in this respect, which start either from Maxwell's equations (see [5, 6] and the review in [7]) or from the dispersion relation for the photon [8, 9], agree that the single-photon wavefunction should be related to the classical electromagnetic field.

The aim of the present paper is to show that the solution of Maxwell's equations can be also regarded as the quantum wavefunction of a single- or of many-photon systems, if the photon is correctly understood as the energy quanta of a monochromatic electromagnetic field. Our approach is based on the association of photons with each spectral and polarization component of the classical electromagnetic field, the scaling procedure that we employ bestowing to the photon wavefunction the interpretation of probability density amplitude, in contrast to the meaning of the Riemann-Silberstein wavefunction as amplitude of the photon energy density. We restrict the discussion in this paper to electromagnetic fields that propagate in vacuum in a certain direction but, unlike other proposals for a photon wavefunction, we consider that the electric and magnetic fields can take complex values. This is a desirable property when phase-related phenomena (such as interference) are referred to. In particular, experimental evidence suggests that in quantum interference with light beams the electric-field operator, instead of the state vector, acquires the geometric phase (see [10] and the references therein), and that it is always the Hannay angle, and not the Berry phase, that is measured in light-beam interference experiments because for light beams the electric-field amplitudes are superposed (in particle interference experiments superposition of wavefunctions takes place). Therefore, there is strong evidence for the need of incorporating complex electric and magnetic fields in a theory of photon wavefunction.

## 2. MAXWELL'S EQUATIONS IN VACUUM

We start from Maxwell's equations in vacuum, written in the form

$$\sqrt{\varepsilon_0}\nabla(\sqrt{\varepsilon_0}\mathbf{E}) = 0 \qquad \nabla\times(\sqrt{\varepsilon_0}\mathbf{E}) = -c^{-1}\partial(\sqrt{\mu_0}\mathbf{H})/\partial t$$
$$\sqrt{\mu_0}\nabla(\sqrt{\mu_0}\mathbf{H}) = 0 \qquad \nabla\times(\sqrt{\mu_0}\mathbf{H}) = c^{-1}\partial(\sqrt{\varepsilon_0}\mathbf{E})/\partial t \qquad (1)$$

to define the wavefunction $\Psi_{em} = \begin{pmatrix} \sqrt{\varepsilon_0}\mathbf{E} \\ \sqrt{\mu_0}\mathbf{H} \end{pmatrix}$, solution of

$$\partial\Psi_{em}/\partial t = c\nabla\times(\mathbf{J}\Psi_{em}) = c(\mathbf{S}\cdot\nabla/i)(\mathbf{J}\Psi_{em}) \qquad (2a)$$
$$\nabla\Psi_{em} = 0 \qquad (2b)$$

with $\mathbf{J} = \begin{pmatrix} 0 & \mathbf{I} \\ -\mathbf{I} & 0 \end{pmatrix}$, $\mathbf{I}$ the 3×3 identity matrix, and $\mathbf{S}$ the spin-1 matrix with components

$$S_x = \begin{pmatrix} 0 & 0 & 0 \\ 0 & 0 & -i \\ 0 & i & 0 \end{pmatrix}, \quad S_y = \begin{pmatrix} 0 & 0 & i \\ 0 & 0 & 0 \\ -i & 0 & 0 \end{pmatrix}, \quad S_z = \begin{pmatrix} 0 & -i & 0 \\ i & 0 & 0 \\ 0 & 0 & 0 \end{pmatrix}. \qquad (3)$$

(To not overcomplicate the notations, the operators $c(\mathbf{S}\cdot\nabla/i)$ in (2a) and $\nabla$ in (2b) are assumed to act separately on the upper and lower components of the vectors at their right-hand-side.) The two components of $\Psi_{em}$ are the electric and magnetic fields, scaled with respect to the squared values of the electric permittivity and magnetic permeability of vacuum, respectively; the subscript *em* indicates that we refer to the classical electromagnetic field. Unlike in the Riemann-Silberstein approach to the photon wavefunction (reviewed in [7]) and in the closely connected analogy between the Dirac equations and Maxwell's equations discovered by Majorana [11], the electric and magnetic fields in (2a,b) can take also complex values.

Maxwell's equations (2a,b) in the Fourier-transform space take the form

$$\partial \Phi_{em}/\partial t = c(\mathbf{S}\cdot\mathbf{k})\mathbf{J}\Phi_{em} \qquad (4a)$$

$$\mathbf{k}\cdot\Phi_{em} = 0 \qquad (4b)$$

where the Fourier-transform of the wavefunction $\Psi_{em}$ is defined as

$$\Phi_{em}(\mathbf{k},t) = (2\pi)^{-3/2}\int \Psi_{em}(\mathbf{r},t)\exp(-i\mathbf{r}\cdot\mathbf{k})d\mathbf{r}, \qquad (5)$$

with $\mathbf{k}$ the wavevector of the classical electromagnetic field. The multiplicative action of the $\mathbf{k}$ wavevector on $\Phi_{em}(\mathbf{k},t)$ is equivalent to the action of the gradient operator $\nabla/i$ on $\Psi_{em}(\mathbf{r},t)$. In this derivation of equations (4a) and (4b) (directly from Maxwell's equations rather than from the second-order d'Alembert equation) there is no ambiguity related to the presence of a scalar quantity on their right-hand-side [12].

In terms of the wavefunction $\Psi_{em}$ the energy flux density (Poynting vector) of the electromagnetic field can be expressed as $\mathbf{j}_{em}(\mathbf{r},t) = i\Psi_{em}^{*}c\mathbf{S}(\mathbf{J}\Psi_{em}) = c(\mathbf{E}^{*}\times\mathbf{H} - \mathbf{H}^{*}\times\mathbf{E})$.

The energy density $\Psi_{em}^*\Psi_{em}$ and $j_{em}$ satisfy the continuity equation $\partial \Psi_{em}^*\Psi_{em}/\partial t + \nabla j_{em} = 0$ (the product of two vectors $a = \begin{pmatrix} a_1 \\ a_2 \end{pmatrix}$, $b = \begin{pmatrix} b_1 \\ b_2 \end{pmatrix}$ is to be understood as $ab = a_1 b_1 + a_2 b_2$).

The monochromatic plane wave solutions of (2a,b), for which $E$, $H$ $\propto \exp(i\mathbf{k}\cdot\mathbf{r} - i\omega t)$, can always be expressed as a superposition of two independent waves, which correspond to the two photon helicities (or polarizations). These can be taken, for example, as the (normalized to unity in the mode continuum limit) waves

$$\Psi_\pm(\mathbf{r},t;\mathbf{k}) = [2(2\pi)^3]^{-1/2}\psi_\pm(\mathbf{k})\exp(i\mathbf{k}\cdot\mathbf{r} - i\omega t), \tag{6}$$

with

$$\psi_\pm(\mathbf{k}) = \begin{pmatrix} f_\pm(\mathbf{k}) \\ \mp i f_\pm(\mathbf{k}) \end{pmatrix} = \begin{pmatrix} f_\pm(\mathbf{k}) \\ \dfrac{\mathbf{k}}{|\mathbf{k}|} \times f_\pm(\mathbf{k}) \end{pmatrix}, \quad f_\pm(\mathbf{k}) = \frac{1}{\sqrt{2|\mathbf{k}|^2 (k_x^2 + k_y^2)}} \begin{pmatrix} -k_x k_z \pm i k_y |\mathbf{k}| \\ -k_y k_z \mp i k_x |\mathbf{k}| \\ k_x^2 + k_y^2 \end{pmatrix}, \tag{7}$$

since $\pm i f_\pm(\mathbf{k}) = -\mathbf{k}\times f_\pm(\mathbf{k})/|\mathbf{k}|$. The solutions (6) were employed also in [13] to construct a complete basis in which the photon state can be expanded; in this section we refer to (6) as polarization states of the classical electromagnetic field. These solutions are ortonormal.

The dispersion relation guarantees that $\omega^2 = c^2|\mathbf{k}|^2$, i.e. that $\omega = \pm c|\mathbf{k}|$. We consider throughout this paper only the solution with positive frequency $\omega$, which describes a forward propagating wave, the negative-$\omega$ solution representing a positive-frequency electromagnetic wave (with opposite helicity) that propagates backward in time [6]; although mathematically this solution exists, there is no hard evidence for its physical reality and, moreover, theoretical considerations show that its existence would contradict experience [14]. Even if they exit, negative-frequency photons are not distinct anti-photon states [15], the anti-particle of the photon being also a photon, and thus carrying the same information about the state of the electromagnetic field.

The complex conjugates of (6) are also monochromatic plane-wave solutions of (2a,b) with a $\exp(-i\boldsymbol{k}\cdot\boldsymbol{r}+i\omega t)$ dependence, which propagate in an opposite direction compared to (6). We disregard for now these solutions since we refer explicitly throughout this paper to electromagnetic waves propagating in one direction; a superposition of waves that propagate in opposite directions leads eventually to stationary states, which do not form the object of our study.

## 3. THE SINGLE-PHOTON WAVEFUNCTION

The point of view that the optical wave is connected with the probability of spatial localization of one photon was explicitly stated in [16]; Dirac argued that the particle and wave properties of light can be reconciled in a quantum theory if one of the optical wave functions is associated to each translational state of the photon. A more quantitative association of the classical electromagnetic field with quantum theory was put forward by Feynman [17], which showed that the classical Maxwell's equations can be obtained from the Newton's equation of motion if the quantum non-commutativity between position and momentum is explicitly accounted for. The identification of the classical electromagnetic field, with real electric and magnetic field components, with the photon energy wavefunction (photon localizability being identified with energy localizability) was subsequently defended by Bialynicki-Birula [15].

Here we state that the classical electromagnetic field in the $\boldsymbol{k}$-space is proportional to the quantum wavefunction of a single photon in the momentum representation. When extended to the coordinate representation this relation becomes identical to the relation between the Landau-Peierls wavefunction and the classical electromagnetic field, criticized in

[7]. The appropriateness of this extension and the related interpretational problem will be discussed in detail in Section 5.

The interpretation of $\Psi_{em}$ as a single-photon wavefunction, $\Psi$ (to avoid confusions, we drop the subscript *em* when we refer to quantum states and variables), originates in the analogy of the Schrödinger equation for massive particles with (2a), if the latter is multiplied on both sides with $i\hbar$:

$$i\hbar \frac{\partial}{\partial t}\Psi = ic(\boldsymbol{S} \cdot \boldsymbol{p})(\boldsymbol{J}\Psi) = H\Psi . \tag{8}$$

The associated Hamiltonian for photon propagation in vacuum is $H = ic(\boldsymbol{S} \cdot \boldsymbol{p})\boldsymbol{J}$, where the momentum operator is now defined as $\boldsymbol{p} = (\hbar/i)\nabla$.

In the quantum theory of radiation, however, the photon is the (stationary) eigenstate of the momentum operator, characterized by the momentum vector $\boldsymbol{p}$ and the polarization. So, in order to define the photon wavefunction one should look at the stationary states of the quantum equation

$$i\hbar \partial\Phi/\partial t = ic(\boldsymbol{S} \cdot \boldsymbol{p})(\boldsymbol{J}\Phi) \tag{9}$$

obtained from (4a) by multiplication with $i\hbar$. Note that we start from a quantum equation obtained directly from the classical Maxwell's equation in the $\boldsymbol{k}$ space and not from the d'Alembert equation as in [8]. In a similar way as in the previous section, the two possible polarization states of the photon for a given $\boldsymbol{p} = \hbar\boldsymbol{k}$ are the normalized wavefunctions

$$\Phi_\pm(\boldsymbol{k},t) = [2(2\pi)^3]^{-1/2}\psi_\pm(\boldsymbol{k})\exp(-i\omega t) \tag{10}$$

8It can be easily checked that for both these states the eigenvalues of the $H = ic(\mathbf{S} \cdot \mathbf{p})J$ and $\mathbf{p}$ operators are, respectively, $\hbar\omega$ and $\hbar\mathbf{k}$, and that $\psi_\pm$ are eigenfunctions of the helicity operator $\hbar \mathbf{S} \cdot \mathbf{p} / |\mathbf{p}|$ with eigenvalues $\pm\hbar$. The relations $\mathbf{p} = \hbar\mathbf{k}$ or $E = \hbar\omega = |\mathbf{p}|c$, with $E$ the positive photon energy, are a consequence of the linearity of Maxwell's equations (which impose the linear dispersion relation $\omega = |\mathbf{k}|c$ for the electromagnetic field) and of the mathematical proportionality between the quantum momentum operator in the position representation and the action of the electromagnetic wavevector. Due to these linear relations, the equality $E/\omega = \mathbf{p}/\mathbf{k}$ holds mathematically for any numerical value of the ratio; however, experiments show that the stationary eigenvalues of the Hamiltonian operator in the Schrödinger-like equation have the meaning of the energy of a quantum particle (photon) only when Maxwell's equation (2a) is multiplied with $i\hbar$. Note that the positive-energy states of photons are the only ones retained in the quantum treatment for the same reason as in the classical theory; unlike in [13], the polarization (helicity) states for classical and quantum theories are treated in the same manner, and negative-energy states are not identified, as in [5, 13] with left-handed polarization in classical theory (versus positive-energy states that describe right-handed polarization).

The general form of the wavefunction in the momentum representation of the photon with a definite wavevector $\mathbf{k}$, which is a non-controversial object, can then be expressed as

$$\Phi(\mathbf{k},t) = [2(2\pi)^3]^{-1/2} [c_{+,\mathbf{k}} \psi_+(\mathbf{k}) + c_{-,\mathbf{k}} \psi_-(\mathbf{k})] \exp(-i\omega t), \quad (11)$$

where the expansion coefficients $c_{\pm,\mathbf{k}}$ depend parametrically on $\mathbf{k}$ because they can be different for different $\mathbf{k}$ values. If $\Phi$ is to be normalized to unity, then $|c_{+,\mathbf{k}}|^2 + |c_{-,\mathbf{k}}|^2 = 1$. Since $\Phi$ satisfies (9), which is the classical Maxwell's equation (4a) multiplied with a constant, it is to be expected that $\Phi$ is proportional to the classical wavefunction $\Phi_{em}$. The constant of proportionality is easily found since for an electromagnetic field with only one photon the total energy should be $\int \Phi^*_{em}(\boldsymbol{k},t)\Phi_{em}(\boldsymbol{k},t)d\boldsymbol{k} = \int \Psi^*_{em}(\boldsymbol{r},t)\Psi_{em}(\boldsymbol{r},t)d\boldsymbol{r}$ $= \int(\varepsilon|\boldsymbol{E}(\boldsymbol{r},t)|^2 + \mu|\boldsymbol{H}(\boldsymbol{r},t)|^2)d\boldsymbol{r} = \hbar\omega$. One can then make the formal identification $\Phi(\boldsymbol{p},t) \equiv \Phi_{em}(\boldsymbol{k},t)/\sqrt{\hbar c|\boldsymbol{k}|}$, relating the photon wavefunction in momentum representation with a scaled classical electromagnetic field with the same wavevector. The scaling factor $(\hbar\omega)^{1/2}$ is in agreement with that in [8], but used in quite a different sense: it is employed here to find the photon wavefunction in momentum representation from the classical electromagnetic field, whereas in [8] it related the photon wavefunction in coordinate representation to the probability amplitudes for photons with given momentum and helicity.

One can then easily check that the quantum photon wavefunction in the coordinate representation, given by $\Psi(\boldsymbol{r},t) = (2\pi\hbar)^{-3/2}\int \Phi(\boldsymbol{p},t)\exp(i\boldsymbol{k}\cdot\boldsymbol{r}/\hbar)d\boldsymbol{p}$, satisfies indeed the quantum equation (8). It follows that, similarly to [13], the photon wavefunction in the coordinate representation can be expressed as

$$\Psi(\boldsymbol{r},t) = [2(2\pi)^3]^{-1/2}\int d\boldsymbol{k}[c_{+,\boldsymbol{k}}\psi_+(\boldsymbol{k}) + c_{-,\boldsymbol{k}}\psi_-(\boldsymbol{k})]\exp(i\boldsymbol{k}\cdot\boldsymbol{r} - i\omega t) \qquad (12)$$

with $\psi_\pm(\boldsymbol{k})$ given by (6). This photon wavefunction is correctly normalized to unity, $\int \Psi^*(\boldsymbol{r},t)\Psi(\boldsymbol{r},t)d\boldsymbol{r} = 1$, and represents the probability density amplitude of photons with all possible frequencies present in the classical electromagnetic field. The photon wavefunction (12) is in agreement with the result in [5], except that we discard negative-energy states as unphysical and employ instead different polarization components. Besides the Hamiltonian $H = ic(\boldsymbol{S}\cdot\hbar\nabla/i)\boldsymbol{J}$ and momentum operator $\boldsymbol{p} = (\hbar/i)\nabla$, which were already identified above, the angular momentum operator is given by $\boldsymbol{r}\times\hbar\nabla/i + \hbar\boldsymbol{S}$.

With this identification of the wavefunction of a single photon, one can introduce, as in [13], a probability current density $j(r,t) = \Psi^* \nabla_p H \Psi = \Psi^* ic S J \Psi = c(E^* \times H - H^* \times E)$, where $\nabla_p$ is the gradient operator in the $p$ space. The photon probability density $\Psi^+ \Psi$ and $j$ satisfy the continuity equation $\partial \Psi^+ \Psi / \partial t + \nabla j = 0$.

## 4. MANY-PHOTON STATES

The many-photon state is an operator obtained from the single-photon wavefunction by replacing the coefficients $c_{\pm,k}$ by operators $a_{\pm,k}$, which describe the annihilation of photons with momentum $p = \hbar k$ and polarization $\pm$. As a result, the many-photon state operator is

$$\Psi(r,t) = [2(2\pi)^3]^{-1/2} \int dk [\psi_+(k) a_{+,k} + \psi_-(k) a_{-,k}] \exp(ik \cdot r - i\omega t), \tag{13}$$

the relation between the many-photon state operator and the positive-frequency part of the electromagnetic field operators obtained in this paper being the same as in [18], although here the presence of only the positive-frequency part for propagating waves is physically justified.

The Hamiltonian operator for a many-photon state is obtained (see [13]) as expectation value of the one-particle energy operator, the result being:

$$H = \int dk \hbar c |k| (a^+_{+,k} a_{+,k} + a^+_{-,k} a_{-,k}). \tag{14}$$

with $a^+_{\pm,k}$ creation operators of photons with momentum $p = \hbar k$ and polarization $\pm$. The total energy in the field at a certain wavevector $k$ is thus a sum over the possible helicities of photons with energies $\hbar \omega = \hbar c |k|$, the number of photons with a given helicity being given by $N_{\pm,k} = a^+_{\pm,k} a_{\pm,k}$. The Hamiltonian does not include the zero-point energy since only propagating electromagnetic fields are considered. As can be seen from a detailed treatment of the quantization of the radiation field [3], the zero-point energy is a consequence of considering stationary electromagnetic waves with real-valued electric and magnetic fields, obtained by superposing counterpropagating waves with complex conjugate amplitudes. Such stationary fields are indeed encountered in situations when matter is in equilibrium with the electromagnetic radiation (when both creation and absorption of photons take place), which explains the appearance of zero-point energy in related phenomena. However, propagating electromagnetic fields cannot be expressed as a superposition of counterpropagating waves, and hence the zero-point energy should not appear in this case. The energy of the many-photon state propagating in one direction (14), differs from the energy of an electromagnetic field with real electric and magnetic field operators, i.e. an electromagnetic wave composed of counterpropagating waves with complex conjugate amplitudes, by exactly the zero-point energy [8], which endorses the interpretation of zero-point energy as originating from cavity-like electromagnetic fields.

The photon state operator (13), with its associated probability current density operator were introduced in [19], the operator that represents the number of photons in a volume $V$ being given by $N_V = \int_V d\mathbf{r} \Psi^+(\mathbf{r},t) \Psi(\mathbf{r},t)$, and the number of photons that cross a surface $\Sigma$ in the temporal interval $(t_1, t_2)$ being defined as $N_\Sigma = \int_{t_1}^{t_2} dt \int_\Sigma dA \hat{\mathbf{n}} \cdot \mathbf{j}(\mathbf{r},t)$, where $\hat{\mathbf{n}}$ is the unit normal to $\Sigma$ in the concerned direction. These two quantities have all the properties of number operators when the linear dimensions of the volume $V$ and surface $\Sigma$ are much larger than the wavelength; they are, in this sense, coarse-grained photon density and photon current density operators. Moreover, the (Maxwell's) equations satisfied by the many-photon state operator were shown to be Lorentz invariant in form [18], although the photon state operator does not transform as a tensor under Lorentz transformations. Another photon number density operator expressed in terms of the electric field and vector potential operators was introduced in [20], which resembles the Mandel photon number operator [21].

## 5. DISCUSSIONS

The photon wavefunction in this paper is in agreement with the approach in [22], which explicitly constructed single-photon wavefunctions defined through $|\Psi\rangle = (2\pi)^{-3} \int d\boldsymbol{k}[g_+(\boldsymbol{k})a^+(+,\boldsymbol{k}) + g_-(\boldsymbol{k})a^+(-,\boldsymbol{k})]|0\rangle$, where the complex-valued functions $g_\pm(\boldsymbol{k})$ are related to the Fourier transform of the vector potential of the classical electromagnetic field, $\boldsymbol{A}(\boldsymbol{k})$, through $g_+(\boldsymbol{k})\hat{\boldsymbol{e}}_+(\boldsymbol{k}) + g_-(\boldsymbol{k})\boldsymbol{e}_-(\boldsymbol{k}) = (\varepsilon_0 \omega/\hbar)^{1/2} \boldsymbol{A}(\boldsymbol{k})$, with $\hat{\boldsymbol{e}}_\pm(\boldsymbol{k})$ unit vectors along the two polarization directions. Indeed, for monochromatic plane-wave solutions of Maxwell's equations both upper and lower components of $\psi_\pm(\boldsymbol{k})$ are proportional to $(\varepsilon_0 \omega/\hbar)^{1/2} \boldsymbol{A}(\boldsymbol{k})$.

The photon wavefunction (12) is formally identical to that in [13], with an important difference: the relation with the classical electromagnetic field (the starting point in [13] is the vector field operators in [19], whereas in our approach the starting point is the classical electromagnetic field) and the justification of neglecting the negative-energy states and the counter-propagating waves. We have retained only the positive-energy states, as in [19], justifying our choice by the propagating nature of the electromagnetic field. The negative-energy states are here regarded as not physical in both quantum and classical theories, unlike in [5] and [13], where it is claimed that negative-energy states are unavoidable in classical electromagnetism. Moreover, [13] disregards the complex conjugates of (6) as unphysical, a point of view that is not adopted here: we disregard them as representing counterpropagating waves.

Note that, in equation (12), because of our assignation of photons to spectral components of the electromagnetic field, no direct relation is established between the photon wavefunction and the classical electromagnetic field as solution of (2a,2b). Such a relation would imply a convolution operation in the spatial coordinates [18], the relation between the photon wavefunction (12) and the classical field being identical to that between Landau-Peierls wavefunction and the electromagnetic field. Because of some drawbacks of the Landau-Peierls functions, Bialynicki-Birula [7] identified $\Psi_{em}$ with the photon energy wavefunction, in the sense that $\Psi_{em}^{*}\Psi_{em}$ represents the energy density. (In fact, the photon wavefunction in [7] is a complex function whose real and imaginary parts are the real parts of the upper and lower components of $\Psi_{em}$; $\Psi_{em}$ can, however, be regarded as another way of expressing the wavefunction in [7] for real electric and magnetic fields). In [7] (and [8]) it is, moreover, argued that no photon wavefunction exists. It seems rather odd that, at least for a monochromatic field, there is photon energy density, but no photon density; in this case the two wavefunctions should be proportional. Another strange feature of the theory in [7] is that different normalization procedures (in general, scalar products) are introduced for wavefunctions in the momentum and position representations. (The treatment for wavefunctions in momentum representation is the same as here.) The normalization procedure in [7] for wavefunctions in coordinate representation is necessary to recover the classical expressions for total and angular momentum of a stationary electromagnetic field from quantum expressions. Besides the fact that a different treatment of coordinate and momentum spaces is not justified, the form of the scalar product for wavefunctions in the coordinate representation, which involves the inverse of the Hamiltonian operator (and hence is interaction- and coordinate-system-dependent), obscures in fact the equivalence of the approach in [7] with the Landau-Peierls-type wavefunctions, as acknowledged in Section 5.3 of [7]. More disturbing, $\Psi_{em}$ can in no way be directly related to photons. It should not be overlooked that the photon is not just the energy quanta of the electromagnetic field, but is the energy quanta of a certain spectral (and polarization) component of the classical field. In this sense, $\Psi_{em}^*\Psi_{em}$ normalized with the average energy [15] cannot represent the photon probability density because the spectral information, and thus the basic ingredient for photon definition, is not explicitly present. The classical quantity $\Psi_{em}^*\Psi_{em}$, when normalized to the total energy, has the meaning of photon density only for stationary, monochromatic fields. In the general case the normalization should not be performed with respect to the total energy but to the energy of a photon at each correspondent frequency of the electromagnetic field, as shown in this paper (a similar argument is to be found in [5]). The possibility to distinguish between photon density and energy density is particularly important when dealing with polychromatic radiation. We can have polychromatic electromagnetic fields, but never polychromatic photons; photons with different energy must be associated with each spectral component.

On the other hand, the nasty mathematical properties of the Landau-Peierls wavefunctions are to be expected from physical considerations. Quoting the criticism of Pauli, Bialynicki-Birula [7] disregards the Landau-Peierls wavefunctions on the reasons that they do not transform as a tensor field (which is true, but still the equations satisfied by them are Lorentz invariant [18]), that they are nonlocal and therefore cannot describe the interaction of the electromagnetic field with localized charges. The fact that these functions are nonlocal is physically predictable since they are a sum of spectral components of the electromagnetic field scaled to the square root of the corresponding photon energy, so that when changing the coordinate system the photon wavefunction at one position must depend on the wavefunction in the original coordinate system at all positions due to the coordinate-system-dependence of frequency. Even for mathematical reasons a nonlocal (physically justified) wavefunction should not be less preferable than a scalar product that does not take the same form for wavefunctions in coordinate and momentum representations and that, in the first case, is dependent on interactions and coordinate systems. The statement that Landau-Peierls wavefunctions are not appropriate to describe the interaction with localized charges is also misleading for two reasons: (i) the interaction of the electromagnetic radiation with matter involves usually waves with quite narrow spectral bandwidth, for which the photon density and photon energy density are almost proportional, and (ii) such interactions are energy conserving and the photon energy density is perhaps a more intuitive wavefunction in these cases than the photon density; there is no real reason why the latter cannot be employed, if carefully applied. The last argument in [7] against the Landau-Peierls wavefunctions is that if one attempts to recover the electromagnetic field from a photon wavefunction that decays abruptly at the boundary, the resulting energy density is infinite. This maybe true mathematically. However, the photon wavefunction in (12) is obtained scaling the spectral components of the classical electromagnetic field at $(\hbar\omega)^{1/2}$. It is hard to imagine that for real electromagnetic fields the situation just described happens. Despite that the wavefunction in (12), which has the same relation to the classical electromagnetic field as the Landau-Peierls wavefunctions, has not all the mathematically desirable properties, it is much more related to photons, as quanta of electromagnetic energy of a certain frequency, than the proposal put forward in [7].

## 6. CONCLUSIONS

We have defined a wavefunction for single- and many-photon states by associating photons with different momenta with different spectral and polarization components of the classical, generally complex, electromagnetic field propagating in a definite direction; a scaling of the latter to the square root of the photon energy provides an appropriately normalized photon wavefunction with the desired interpretation of probability density amplitude. In this respect, the photon wavefunction introduced in this paper is completely different (in coordinate but not in the momentum representation) from the photon energy wavefunction in [7]. The differences between these two wavefunctions have been discussed in detail in Section 5. Our approach resembles to that in [5], except for a different interpretation of negative-energy states and for the extension to many-photon states performed in this work. It also parallels the introduction of the wavefunction in [13], but based on a completely different point of view: the photon wavefunction in this paper is directly related to the classical electromagnetic field, whereas in [13] the starting point are vector field operators previously used in [18, 19] to define a coarse-grained photon density and photon current density operators. The impossibility of defining a photon density in the coordinate representation as the projection of the state vector on the eigenstates of the position operator was demonstrated in [1], where it was shown that a position operator for photons cannot be defined. Reference [1], however, attempts to define localized states of an elementary system, understood as elementary systems localized at the space-time origin ($r = 0$, $t = 0$), from the demands that all states of the system can be obtained from superpositions of relativistic transforms of any state. According to this criterion, which is not even relativistic invariant for massive particles (see the discussion in [7]), and in contradiction to experiments, photons appear to be not localizable [1]. Because photons propagate with velocity $c$ and therefore the above definition of localized states is not appropriate for them (and hence, the conclusion that photons cannot be localized in the above sense is irrelevant), nothing prevents photons to be coarse-grained localizable and to have an associated photon wavefunction that is not necessarily related to state projections on the position operator.